%%%%    This Paper Starts from Here!!!    %%%%%%%%%%%%%%
%%%%%%%%%%%%%%%%%%%%%%%%%%%%%%%%%%%%%%%%%%%%%%%%%%%%%%%%
\magnification=1200
\input psfig.tex
\hsize=6.3 truein
\vsize=8.4 truein

\baselineskip 20 pt
\centerline{\bf Hyperspherical Close-Coupling Calculation of D-wave 
 Positronium Formation}
\centerline{\bf  and Excitation Cross Sections 
 in Positron-Hydrogen
Scattering}
\vskip .5truein
\centerline{Yan Zhou and  C. D. Lin$^a$ }
\medskip
\centerline{Department of Physics, Kansas State University}
\centerline{Manhattan, KS 66506}
\medskip
\centerline{$^a$ JILA visiting Fellow. JILA, Box 440, University of Colorado}
\centerline{Boulder, Co. 80309}
\vskip .3 truein
\centerline{\bf Abstract}
\bigskip

Hyperspherical close-coupling method is used to calculate the elastic,
positronium formation and excitation cross sections for
positron collisions with atomic
hydrogen at energies below the H(n=4) threshold for the J=2  partial wave.
The resonances below each inelastic threshold are also analyzed. The adiabatic
hyperspherical potential curves are used to identify the nature of these
resonances.
\bigskip

{\bf 1. Introduction}

   The scattering between a positron and an atomic hydrogen is one of the 
simplest-looking quantum mechanical three-body problems. Yet it poses major
challenge for theorists as well as for experimentalists over the years. The 
difficulty of formulating a satisfactory theoretical approach which accounts
for all the possible inelastic transitions, including excitation, positronium formation,
and ionization processes in a broad energy range is responsible for the slow
progress in this field.  

   In the past, positron-hydrogen atom scattering has been treated theoretically
using the standard methods generalized from electron-atom scattering, 
such as the close
coupling method [1-6], the R-matrix method [7], and the variational method
[8-9].
 There
are some successes, especially in the lower energy region where few channels are open.
As the collision energy increases the number of open
inelastic channels increases 
rapidly and there are few results in the higher energy region where all the 
possible final channels are addressed.

  An alternative approach, developed in the last few years for treating
positron-hydrogen atom scattering, is the hyperspherical close coupling
(HSCC) method. The HSCC has been initially developed for two-electron
atomic systems [10], but has since been generalized to any three-body
systems recently by Igarashi and Toshima [11], and by Zhou and Lin [12].     
The basic formulation is similar to the earlier work of
Archer $et\ al$ [13] but the  
recent success of the HSCC method  owes much to the improvement in
the numerical 
accuracy which makes the detailed studies
of this collision system possible.

  The HSCC method has been used to study in details the J=0 and J=1
partial waves in e$^+$+ H collisions in the energy region below 
the H(n=4) threshold [14]. Partial cross sections to each final state
and the resonances below and near each inelastic threshold were 
examined. There are few inelastic cross sections available for 
comparison in the higher energy region except some recent ones
from the close coupling method [3]. For the resonance parameters, 
calculations based on the complex coordinate rotation method [15,16]
have been carried out for a number of states and they compare
well with the results from the HSCC method [14]. The HSCC has also 
been used to show that there are no resonances in the positronium
formation channel at energies above the ionization threshold [17]. This
is a subject of great controversies [4,6,11,18-21]. The HSCC method also has been
applied to calculate and compare resonances in the collision systems:
$e^-$+H, $e^+$+H and $e^-$+Ps [22]. In carrying out these calculations, the
same set of codes are used since the HSCC method has been developed for
the general three-body systems.

  In this paper we report new results for the J=2 partial wave for
the positron-hydrogen atom scattering using the HSCC method. The HSCC
method is briefly reviewed in Section 2 and recent numerical
implementations  
are addressed. The results for the 
calculated inelastic scattering cross sections and resonance
parameters are presented in Section 3. Adiabatic hyperspherical
potential curves are used to assist the interpretation of the 
calculated resonances. A final summary is given in Section 4. 
\bigskip
{\bf 2. The Hyperspherical Close Coupling Method}
 
  First we define the Jacobi coordinates of the three particles. There are
three possible choices of the Jacobi coordinates, as depicted in Fig. 1.
From each set of Jacobi coordinates, $\vec\rho_1$ and $\vec\rho_2$, a set of
hyperspherical coordinates can be defined. In hyperspherical
coordinates, the wavefunctions $\Psi$ 
are shown to satisfy
$$(-{\partial^2\over \partial R^2}+H_{ad}-2\mu E)\Psi(R,\phi,\hat\Omega)=0,\eqno(1
)$$
where R is the hyperradius, $\phi$ is the hyperangle and 
$\hat\Omega$ denotes collectively the four orientation angles of vectors
$\vec\rho_1$ and $\vec\rho_2$. The adiabatic Hamiltonian $H_{ad}$ is the
total Hamiltonian of the system in the center-of-mass frame evaluated at
a constant value of the hyperradius and $\mu$ is a scaled mass.

  In the HSCC approach, the hyperradius is divided into two regions
at R=$R_0$. In the inner region R$\le R_0$ where all the
three particles interact with each other, the wavefunctions are to be
expressed in terms of hyperspherical coordinates. In the outer region, the
three-body system breaks into a single particle and a pair of particles in some
bound states. In this outer region, the wave functions are appropriately
expressed in terms of Jacobi coordinates for each arrangement.

  In the so-called diabatic-by-sector method, the inner region is further
divided into many small sectors. Within each sector, the wave function
is expanded as 

$$\Psi (R,\phi,\hat\Omega)=\sum_\nu\sum_IF_{\nu I}(R)\Phi_{\nu I
}(R_a;\theta,\phi)\tilde{D}^J_{IM_J}(\omega_1,\omega_2,\omega_3),\eqno(2) $$
where we have expressed the wave functions in the body-frame of the 
system, with the z'-axis chosen to be along the line connecting from the 
positron to the proton. The y'-axis is chosen to be perpendicular to the
plane of the three particles and that the electron is always on the +x'
side. In (2), the normalized D-function has good-parity 
[23], with the $\omega$'s being the Euler angles. The angle $\theta$ is 
between the two vectors,  $\vec\rho_1$ and $\vec\rho_2$,  $\phi$ is
the hyperangle and  
$\nu$ is the channel index. $R_a$ is chosen at the midpoint of the sector,
J is the total angular momentum, I is
the absolute value of the projection of
$\vec J$ along the body frame's $z^\prime$ axis and running from 0 to J for $(-1)^JP
=1$ states and
from 1 to J for $(-1)^JP=-1$ states, with P being the parity. 

   We solve the basis functions $\Phi_{\nu I}(R_a;\theta,\phi)$ at $R_a$
for a fixed I component on the $(\theta,\phi)$ plane using the
finite-element method. The hyperradial function satisfies 
a set
of coupled differential equations 
$$\left(-{\partial^{2}\over\partial R^{2}}-{1\over 4R^2}-2\mu E\right)F_{\mu
I}(R)+\sum_{\nu I^\prime}V_{\mu I,\nu I^\prime}(R)F_{\nu I^\prime}(R)=0\eqno(3) $$
where the coupling matrix elements V's are
defined explicitly in Zhou and Lin [12]. 

   In the practical implementation of the HSCC method, the set of
hyperradial equations (3) are integrated within each sector.
Starting with the innermost sector, the integration is continued
until it reaches the boundary of the next sector. At this boundary,
the total wave functions are expanded in terms of basis functions
of the next sector from which integration within the next sector
can be carried out. This procedure is
continued from small hyperradius to $R_0$
where the wave function is matched to an outside solution expressed in terms
of independent electronic coordinates ${\vec\rho_1, \vec\rho_2}$. 
From the matching, one can extract the K-matrix and 
the partial scattering cross sections

$$\sigma_{ij}^{(J)}={4\pi(2J+1)\over k^2}\left |{K\over 1-iK}\right |^2_{ij} \eqno
(4)$$
where k is the momentum of the incident particle.

 In order to be able to calculate excitation and positronium formation
cross sections to excited states, the matching radius $R_0$ should be
chosen at a relative large value, and the hyperspherical basis functions
also have to be evaluated at large values of R. In the large R region,
the basis functions for the lower channels are located near the singularities
of the potential surfaces. We improve the numerical accuracy by 
introducing two new angles which are more appropriate for describing
the large R region, as discussed in Zhou and Lin [14]. Another 
complication due to the degenerate hydrogenic excited states is that
the coupling among the different subshells in a given principal 
quantum number is quite large. To  accommodate the interaction
among the degenerate hydrogenic states, we use the dipole states of 
Gailitis and Damburg[24] and of Seaton[25]. The details of this transformation
is also discussed in Ref. [14].
\bigskip 
{\bf 3. Results and Discussion}

  To obtain results for D wave scattering, the basis functions for the 
I=0, 1 and 2 components
are first calculated. The matching radius is chosen at $R_0$=250.655 a.u.
and 351 sectors are used in the inner region. Since we are interested 
in energies below
the H(n=4) threshold, in the solution of hyperradial equations (3) we
include 21 basis functions for I=0, 13 for I=1 and 9 for I=2. The basis
set can be enlarged to achieve higher accuracy but at the expense of
more computing time. 

  In presenting the results, we first mention that the thresholds,
in increasing order, for 
Ps(n=1), H(n=2), Ps(n=2), H(n=3) and H(n=4) are located at 0.5, 0.75, 0.875,
0.8889 and 0.9375 Ryd, respectively, above the ground state of H. There 
are resonances below each threshold.

{\bf 3.1. Elastic and Ps(n=1) formation cross sections}

  In Fig. 2 we show the D-wave elastic and Ps(n=1) formation cross sections
for positron scattering with atomic hydrogen in the energy region between 0.7 
and 0.94 Ryd. Both cross sections change little in the whole energy region
shown. There are resonance structures below each inelastic threshold, but 
pronounced ones are observed only below the H(n=2) threshold.  
In the figure we also show the close coupling results of Mitroy and 
Ratnavelu[3]
where they performed close coupling calculations using six atomic states
on each center. Their calculation will be denoted as CC(6,6).
The results from the two calculations are in quite 
reasonable agreement. 

  The agreement in the elastic and Ps(n=1) formation cross sections between
the two calculations, however, does not imply that the predicted 
resonance positions are identical. Both have calculated one Feshbach
resonance below the H(n=2) threshold. The position and the width of 
this resonance are listed in Table 1. Note that the CC(6,6) results
are quite different from the HSCC results. In similar comparison for 
the S- and P-waves results, the CC(6,6) results also show large discrepancies
from the HSCC results in the resonance parameters. In those cases, the 
HSCC results are in good agreement with those obtained variationally
using the complex
coordinate rotation method[15,16]. The CC(6,6) results were different. 
We tend to believe that our results
here are more accurate than those from the CC(6,6) calculations.

  The HSCC method allows one to analyze qualitatively and semiquantitatively
the resonance structures. In Fig. 3 we show the three adiabatic potential curves
that converge to the H(n=2) threshold. Only one curve is attractive
which can support Feshbach resonances. The position of the calculated
resonance is shown as dotted lines in the figure. Note that one would 
expect an infinite series of resonances associated with the lowest curve
since this potential approaches the threshold as ${-0.81/R^2}$.
Such a dipole potential can support an infinite number of resonances, but
all the higher ones are very close to the threshold and were not examined.

{\bf 3.2. Excitation cross sections to the 2s and 2p states}
 
  We show in Fig. 4 the D-wave excitation cross sections to 2s and 2p
states. The results from the CC(6,6) calculations are also shown. Note
that the results from the two calculations are again quite close,
except in the resonance region. We have examined three resonances
below the Ps(n=2) threshold. The positions and the widths are listed
in Table 1 and the results for the first two are compared to the
CC(6,6) results. Again the discrepancy is quite significant.

  Our calculations show small structures near the H(n=2) threshold.  
We suspect that these structures are due to the relatively small 
basis functions used in this calculation. 

   We also show in Fig. 5 the three adiabatic potential curves that converge
to the Ps(n=2) threshold. Note that the lowest curve is rather 
attractive. In fact, it behaves as $-17.49/R^2$ asymptotically.
Note that the second curve in Fig. 5 has a small attractive part in the
inner region, but the potential well is not strong enough to support a
resonance.

{\bf 3.3. Ps(n=2) formation cross sections}

   The cross sections for the positronium formation cross sections to 2s and
2p states are shown in Fig. 6. These cross sections are quite small, about
0.5\% of the cross sections for the elastic and the Ps(n=1) cross sections,
and about 10\% of the n=2 excitation cross sections. The resonances are
quite pronounced in these small channels. The resonance positions and
widths below the H(n=3) threshold are studied and the results are shown
in Table 1. There are no other calculations available for comparison
in this energy range.

{\bf 3.4. Excitation cross sections to n=3 states}

  The excitation cross sections to H(n=3) states have also been calculated.
The results are shown in Fig. 7. These cross sections are about a factor
of two smaller than the Ps(n=2) cross sections in the same energy 
region. The resonances are quite pronounced at energies below the H(n=4)
threshold and the positions and widths of the five lowest resonances
are shown in Table 1. 

   There are no other calculations available for comparison. The small
structures near the threshold are likely due to the small number
of hyperspherical channels used in the present calculation.
 
  Careful analysis of the resonances at the higher thresholds is more
difficult due to the numerous avoided crossings. For example, we show
in Fig. 8 some of the adiabatic curves that converge to the H(n=3)
threshold. We also show that the uppermost curve which converges to 
the Ps(n=2)
limit has many avoided crossings with the H(n=3) curves.
\bigskip
{\bf 4. Summary}

  In this paper we present D-wave results for the excitation and 
positronium formation cross sections in positron-hydrogen atom
collisions at energies below the H(n=4) threshold using the hyperspherical
close coupling method. There are few calculations available for comparison
in the higher energy region considered in this paper. Together
with  our previous
results for the S- and P-wave, we illustrated that the hyperspherical 
close coupling method provides a very powerful method for treating
general rearrangement collisions. Calculations including the 
ionization channels are underway which would allow the HSCC method to
be extended to the higher energy region. The method can be generalized
to collisions between positrons and alkali atoms by employing a
model potential description of the target atom.
\bigskip
{\bf Acknowlegements}

 This work is supported in part by the U.S. Department of Energy, Office of
Energy Research, Office of Basic Energy Sciences, Division of Chemical
Sciences. 
\bigskip

{\bf References}

1. R.N. Hewitt, C.J. Noble and B.H. Bransden, J. Phys. B{\bf 23},4185 (1990).

2. J. Mitroy, J. Phys. B{\bf 26}, 4861 (1993).

3. J. Mitroy and K. Ratnavelu, J. Phys. B{\bf 28}, 287 (1995).

4. A.A. Kernohan, M.T. McAlinden and H.R. Walters, J. Phys. B{\bf
27}, L625 (1994).

5. N.K. Sarkar, M. Basu and A.S. Ghosh, J. Phys. B{\bf 26}, L79 (1993).

6. T.T. Gien and G.G. Liu, J. Phys. B{\bf 27}, L179 (1994) 

7. C.J. Brown and J.W. Humberston, J. Phys. B{\bf 18} L401,
(1985).

8. J.W. Humberston, Adv. At. Mol. Phys. Vol.{\bf 22}, 1 (1986).

9. K. Higgins and P.G. Burke, J. Phys. B{\bf 24}, L343 (1991).

10. J.Z. Tang, S. Watanabe and M. Matsuzawa, Phys. Rev. A{\bf 46}, 2437
(1992).

11. A. Igarashi and N. Toshima, Phys. Rev. A{\bf 50}, 232 (1994).

12. Y. Zhou and C.D. Lin, J. Phys. B{\bf 27}, 5065 (1994).

13. B.J. Archer, G.A. Parker and R.T. Pack, Phys. Rev. A{\bf 41}, 1303 (1990).

14. Y. Zhou and C.D. Lin, submitted to J. Phys. B (1995).

15. Y.K. Ho, J. Phys. B {\bf 23}, L41 (1990).

16. Y.K. Ho, Hyperfine Interactions {\bf 73}, 109 (1992).

17. Y. Zhou and C.D. Lin, submitted to J. Phys. B (1995).

18. R.N. Hewitt, C.J. Noble and B.H. Brandsden, J. Phys. B{\bf 24}, L635
(1990).

19. K. Higgins and P.G. Burke, J. Phys. B{\bf 26}, 4269 (1993).

20. J. Mitroy and A.T. Stelbovics, J. Phys. B{\bf 27}, L55 (1994).

21. T.T. Gien, J. Phys. B{\bf 27}, L25 (1994).

22. Y. Zhou and C.D. Lin. submitted to Phys. Rev. Let. (1995).

23. A.K. Bhatia and A. Temkin, Rev. Mod. Phys. {\bf 36}, 1050 (1964).

24. M. Gailitis and R. Damburg, Proc. Phys. Soc. {\bf 82}, 192 (1963).

25. M.J. Seaton, Proc. Phys. Soc. {\bf 77}, 174 (1961).

\vfill
\eject
\noindent {\bf Table 1}. Positions and widths of D-wave resonances below
different inelastic thresholds for the $e^+$+H collision system. The
present
results are from the HSCC calculation, while the CC(6,6) results 
are from the close coupling calculation of Mitroy and Ratnavelu [3].
The energies and half widths are in units of Rydbergs.

\noindent \vrule height 0.8pt width 5.20in
\vskip -8 pt
\settabs 4 \columns
\+&present&~~~~~~~~~~~~~~~~~~~CC(6,6)&           \cr
\+~~~~~~~~~~~~~~~~~~~E(Ryd)&~~~~~~~~~${1\over2}
\Gamma$(Ryd)&~~~~~~~~~~~E(Ryd)&${1\over 2}\Gamma$(Ryd)\cr
\vskip -8 pt
\noindent \vrule height 0.4pt width 5.20in
\vskip 0.05 truein
\+&~~~~~~~~~~~~Below H(N=2) Threshold&&\cr
\vskip 0.05 truein
\+~~~~~~~~~~~~~~~~~~-0.250031&~~~~~~~~~~1.65(-6)&~~~~~~~~~~-0.25006&~3.9(-4)\cr
\vskip 0.05 truein
\+&~~~~~~~~~~~~Below Ps(N=2) Threshold&&\cr
\vskip 0.05 truein
\+~~~~~~~~~~~~~~~~~~-0.143429&~~~~~~~~~~1.66(-4)&~~~~~~~~~~-0.13715&~1.25(-4)\cr
\+~~~~~~~~~~~~~~~~~~-0.128529&~~~~~~~~~~6.66(-5)&~~~~~~~~~~-0.12765&~3.3(-4)\cr
\+~~~~~~~~~~~~~~~~~~-0.12503&~~~~~~~~~~6.32(-6)&&\cr
\vskip 0.05 truein
\+&~~~~~~~~~~~~Below H(N=3) Threshold&&\cr
\vskip 0.05 truein
\+~~~~~~~~~~~~~~~~~~-0.114529&~~~~~~~~~~4.79(-4)&&\cr
\+~~~~~~~~~~~~~~~~~~-0.111630&~~~~~~~~~~7.89(-5)&&\cr
\vskip 0.05 truein
\+&~~~~~~~~~~~~Below H(N=4) Threshold&&\cr
\vskip 0.05 truein
\+~~~~~~~~~~~~~~~~~~-0.075582&~~~~~~~~~~6.27(-5)&&\cr
\+~~~~~~~~~~~~~~~~~~-0.069775&~~~~~~~~~~6.76(-5)&&\cr
\+~~~~~~~~~~~~~~~~~~-0.066940&~~~~~~~~~~4.70(-5)&&\cr
\+~~~~~~~~~~~~~~~~~~-0.064103&~~~~~~~~~~1.39(-5)&&\cr
\+~~~~~~~~~~~~~~~~~~-0.063855&~~~~~~~~~~2.74(-5)&&\cr
\noindent \vrule height 0.8pt width 5.20in
\vfill
\eject
{\bf Figure Captions} 

Fig. 1. Three sets of Jacobi coordinates for the three particle system.

Fig. 2. D-wave elastic and Ps(n=1) formation cross sections for positron
 scattering with atomic hydrogen. Elastic cross sections: solid lines,
 present results; dotted lines, from CC(6,6)[3].
 Ps(n=1) formation cross sections: long dashed lines, present results;
 dash-dotted lines, CC(6,6). The two results for the Ps(n=1)
 formation are almost indistinguishable. The sharp structures are
 due to the resonances below each inelastic threshold.

Fig. 3. D-wave adiabatic potential curves that converge to the 
    H(n=2) threshold. The position of the lowest Feshbach resonance is
    indicated by the dotted lines.
 
Fig. 4. D-wave excitation cross sections to 2s and 2p states for positron
 scattering with atomic hydrogen. Excitation to 2s: solid line,
 present result; dotted lines, CC(6,6)[3]. Excitation
 to 2p: long dashed lines, present results; dash-dotted lines, CC(6,6).

Fig. 5. D-wave adiabatic potential curves that converge to the
    Ps(n=2) threshold. The position of the first Feshbach resonance is
    indicated by the dotted lines.

Fig. 6. D-wave positronium formation
 cross sections to 2s and 2p states for positron
 scattering with atomic hydrogen obtained from the present calculation
 : solid line for Ps(2s) and dotted lines for Ps(2p).

Fig. 7. D-wave excitation cross sections to H(3s), H(3p) and H(3d) states for
positron scattering with atomic hydrogen. Solid line: excitation to 3s;
dotted lines: excitation to 3p; dashed lines: excitation to 3d.

Fig. 8. D-wave adiabatic potential curves that converge to the
     H(n=3) threshold. The uppermost curve converging to Ps(n=2) is also 
     shown.
\vfill\eject
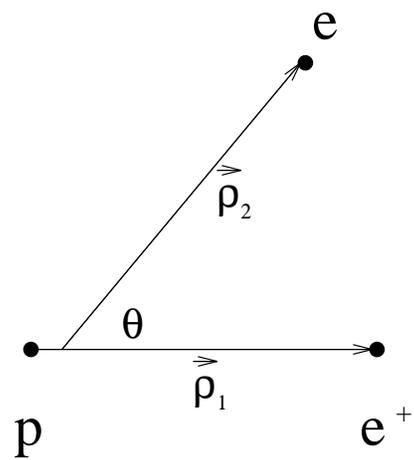
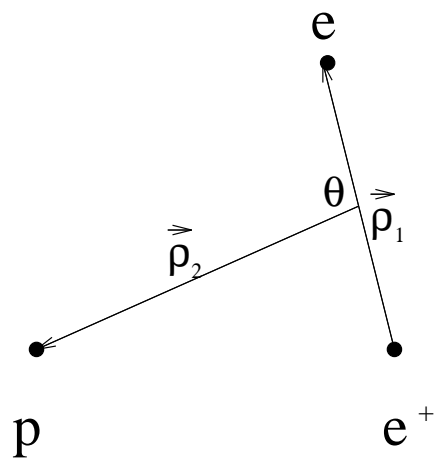
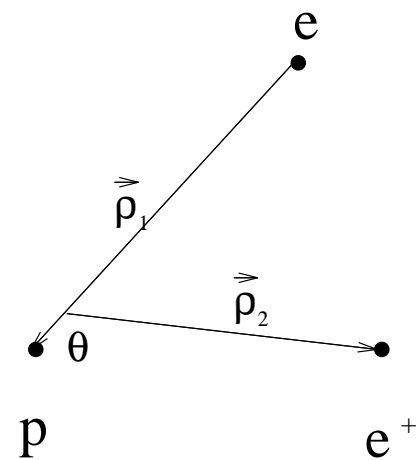
\bye